\documentclass[prb,twocolumn,showpacs,superscriptaddress,floatfix]{revtex4}  
\usepackage{graphicx}\usepackage[dvips]{epsfig}
\usepackage{amssymb,amsmath,latexsym,bm}

\renewcommand{\vec}[1]{{\bm{\mathrm{#1}}}}
\newcommand{\vhat}[1]{\hat{\bm{\mathrm{#1}}}}

\begin{document}
\title{Intrinsic spin torque without spin-orbit coupling}
\author{Kyoung-Whan Kim}%
\affiliation{Basic Science Research Institute, Pohang University of Science
and Technology, Pohang 790-784, Korea}%
\affiliation{PCTP and Department of Physics, Pohang University of Science and
Technology, Pohang 790-784, Korea}%
\affiliation{Center for Nanoscale Science and Technology, National Institute
of Standards and Technology, Gaithersburg, Maryland 20899, USA}%
\affiliation{Maryland NanoCenter, University of Maryland, College Park, Maryland 20742, USA}%
\author{Kyung-Jin Lee}%
\affiliation{Department of Materials Science and Engineering, Korea
University, Seoul 136-701, Korea}%
\affiliation{KU-KIST Graduate School of Converging Science and Technology,
Korea University, Seoul 136-713, Korea}%
\author{Hyun-Woo Lee}%
\email{hwl@postech.ac.kr}%
\affiliation{PCTP and Department of Physics, Pohang University of Science and
Technology, Pohang 790-784, Korea}%
\author{M. D. Stiles}%
\email{mark.stiles@nist.gov}%
\affiliation{Center for Nanoscale Science and Technology, National Institute
of Standards and Technology, Gaithersburg, Maryland 20899, USA}%
\date{\today}

\begin{abstract}
  We derive an intrinsic contribution to
  the non-adiabatic spin torque for non-uniform magnetic textures. It
  differs from previously considered contributions in several ways and
  can be the dominant contribution in some models.  It
  does not depend on the change in occupation of the electron states due to
  the current flow but rather is due to the perturbation of the electronic
  states when an electric field is applied. Therefore it should be viewed
  as electric-field-induced rather than current-induced. Unlike previously
  reported non-adiabatic spin torques, it does not originate from extrinsic
  relaxation mechanisms nor spin-orbit coupling. This intrinsic
  non-adiabatic spin torque is related by a chiral connection to the
  intrinsic spin-orbit torque that has been calculated from the Berry
  phase for Rashba systems.
\end{abstract}


\maketitle

\section{introduction}
Electrical manipulation of magnetization is a promising technique for
enabling a new generation of magnetoelectronic devices. Spin-transfer
torque\cite{Berger84JAP,Slonczewski96JMMM,Berger96PRB,Ralph08JMMM} is an
efficient way to implement the electrical control of magnetization, as has
been demonstrated for various magnetic nanostructures such as spin valves,
magnetic tunnel junctions, and magnetic nanowires.
In the standard picture of spin-transfer
torque, an external electric field generates a spin-polarized electrical
current, which in turn gives rise to \emph{current-induced} spin-transfer
torque. In magnetic nanowires with continuously varying magnetic textures,
this picture leads to two components of current-induced spin torque, which
are known as adiabatic spin torque\cite{Berger84JAP,Tatara04PRL} and
non-adiabatic spin torque.\cite{Zhang04PRL,Thiaville05EPL} The adiabatic
spin torque arises from spin angular momentum conservation when conduction
electron spins adiabatically follow the local magnetization
direction.

The non-adiabatic spin torque, which is perpendicular to the adiabatic spin
torque, arises from a variety of mechanisms and is a crucial factor for
efficient electrical manipulation of magnetic textures such as magnetic
domain walls and skyrmions. One mechanism for non-adiabatic spin torques
occurs only for very short length scale variations in the magnetic
texture,\cite{Tatara07JPSJ,Tatara04PRL,Xiao06PRB} when the spins cannot
adiabatically follow the magnetization texture. In slowly varying magnetic
textures, all previously considered mechanisms for non-adiabatic spin torques
derive from either spin relaxation\cite{Zhang04PRL} or spin-orbit
coupling\cite{Garate09PRB} related to magnetic damping.\cite{Kambersky07PRB}
Here, we describe an intrinsic contribution to the non-adiabatic spin torque
that arises in the slowly varying limit from an effective spin-orbit coupling
due to the magnetic texture.  It is distinguished from other contributions in
that it is \emph{electric-field-induced} rather than current-induced.

The distinction we are trying to draw between electric-field-induced
and current-induced torques is potentially confusing because current
and electric field are proportional to each other.  In linear
response, either torque can be written as proportional to either the
current or the field.  The difference we would like to draw is in how
the leading order constants of proportionality depend on the electron
momentum-relaxation lifetime.  By current-induced torque, we mean one
that is proportional to the current with a coefficient that is
independent of the lifetime and is proportional to the electric field
with a coefficient that is proportional to the lifetime (or
conductivity). By electric-field induced effect, we mean one that is
proportional to the electric field with a coefficient that is
independent of the lifetime and is proportional to the current with a
coefficient that is inversely proportional to the lifetime.

Electric-field-induced spin-transfer torques differ from current-induced
spin-transfer torques in that they do not originate from the electron
occupation change giving rise to current flow. Instead, they originate
from the
perturbation of the electronic states by an external electric field. In general,
electric-field-induced effects depend on the modification of the electron
states summed over the whole Fermi sea, much as
densities involve the sum over all occupied states, while current-induced effects depend on properties only at the
Fermi surface, much as electrical currents do. Examples of electric-field-induced
effects include
voltage-induced magnetic
anisotropy changes,\cite{Wang12NM,Shiota12NM} the
intrinsic spin Hall effect,\cite{Tanaka08PRB} and the intrinsic
spin-orbit torque.\cite{Kurebayashi14NN} Electric-field-induced
torques
are promising for significantly enhancing electrical manipulation
efficiencies.\cite{Shiota12NM,Wang12NM,Kurebayashi14NN,Liu12Science}
Unfortunately their mechanisms are less well understood than current-induced
spin-transfer torques.

In this paper, we examine electron transport through continuously varying
magnetic textures and demonstrate the existence of an electric-field-induced
spin torque. The result is \emph{intrinsic} in the sense that it is
independent of impurity scattering rates. For a free electron dispersion, we
find that this electric-field-induced torque has the same form as the
non-adiabatic spin torque but does not originate from extrinsic relaxation
mechanisms, spin-orbit coupling, nor rapidly varying textures. Moreover, we
demonstrate that it is significantly larger than other contributions to the
non-adiabatic spin torque in some models, making it potentially important for
optimizing the manipulation of magnetic structures such as magnetic domain
walls and Skyrmions.

The intrinsic non-adiabatic spin torque that we report here is closely
related to the intrinsic spin-orbit torque\cite{Kurebayashi14NN} calculated
from a Berry phase.  Previously, we reported\cite{Kim13PRL} that spin-orbit
coupling generates chirality in magnetic properties and that many properties
of a system acquire chiral counterparts upon the introduction of spin-orbit
coupling. We demonstrate below that the intrinsic spin-orbit torque is the
chiral counterpart of the intrinsic non-adiabatic spin torque that we report
here.  This connection indicates the common origin of the two, which can be
computed through a variety of techniques including a Berry phase as done
earlier\cite{Kurebayashi14NN} or perturbation theory like we do here. This
intrinsic spin-orbit torque is also electric-field-induced in the terminology
we use in this paper.

We present our result with a free electron model with exchange splitting for
illustration, but the result can be easily generalized for arbitrary
dispersions. As is the case for the spin Hall effect in the closely related
system with Rashba spin-orbit coupling,\cite{Sinova04PRL,Inoue04PRB} the
intrinsic non-adiabatic spin torque is exactly canceled by vertex corrections
due to spin-independent scattering.\cite{Tatara07JPSJ} However, we
demonstrate that such exact cancellation only occurs for non-magnetic
scatterers\cite{Inoue06PRL} and this particular free-electron model.

This paper is organized as follows. In Sec.~\ref{Sec:Result}, we present our
model and summarize the central results. In Sec.~\ref{Sec:Derivation}, we
provide detailed derivation and some remarks for more motivated readers. In
Sec.~\ref{Sec:Discussion}, we discuss implications of our result, as an
intrinsic origin of non-adiabatic spin-transfer torque. In addition, we
discuss the relationship of these results through Onsager
reciprocity and a chiral connection with previously developed
results. We summarize the paper in Sec.~\ref{Sec:Summary}.

\section{Results\label{Sec:Result}}
In this section we illustrate the results of our calculation by applying it
to a model based on the free electron dispersion and ignore the vertex
corrections.  This model allows us to summarize our key results and provide a
more intuitive description before presenting a formal derivation.  A
derivation and discussion of more general models are given in
Sec.~\ref{Sec:Derivation}.

\begin{figure}
\includegraphics[width=8.6cm]{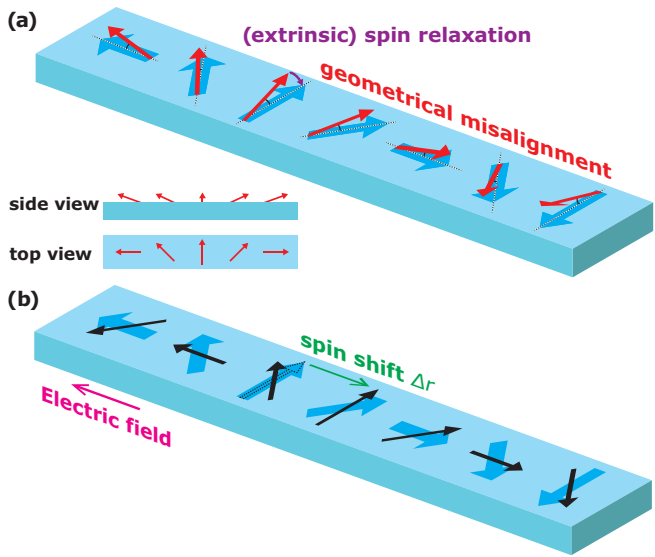}
\caption{(color online) Illustration of electron spin eigenstates in a spin spiral.
(a) The conduction electron spin profile when $\vec{m}$ forms an in-plane
spiral (blue thick arrows) in a magnetic nanowire. When the electric field is absent,
electron spins (red arrows) have a small out-of-plane
component $(\hbar/2J)\vec{m}\times(\vec{v}_{\vec{k}}\cdot\nabla)\vec{m}$ in addition
to the local magnetization direction [Eq.~(\ref{Eq:Spin density (unperturbed, lab)})].
In equilibrium, the out-of-plane deviation from electrons with
momentum $\vec{k}$ and $-\vec{k}$ exactly cancels each other
(see Appendix.~\ref{Sec(A):spin spiral} or Refs.~\onlinecite{Aharonov92PRL,Xiao06PRB}).
However, an electric field changes the occupations (electrical current) and removes the
exact cancellation. The surviving part gives an in-plane spin torque, the
adiabatic spin torque. Extrinsic relaxation (purple arrow) of the
out-of-plane spin deviation gives
an out-of-plane spin torque, the current-induced (or extrinsic) non-adiabatic spin torque. (b)
When an electric field is applied, it not only changes the occupation
of the electron states but also perturbs the eigenstates giving an
additional contribution to spin deviation
in the plane (black arrows) by the spin shift
[Eq.~(\ref{Eq:Spin density (perturbed, lab)})] (green arrow).  This deviation
gives rise to an out-of-plane torque, the electric-field-induced (or intrinsic)
non-adiabatic spin torque. This in-plane deviation does not cancel for
electrons with
momentum $\vec{k}$ and $-\vec{k}$.
In this figure $\alpha_{\vec{k}}$ and $\Delta\vec{r}$ are exaggerated for
clarity.} \label{Fig:Spin Spiral}
\end{figure}

We consider the Hamiltonian
\begin{equation}
H=\frac{\vec{p}^2}{2m_{\rm e}}+J\vec{\sigma}\cdot\vec{m}(\vec{r},t),\label{Eq:s-d Hamiltonian}
\end{equation}
where $\vec{p}$ is the electron momentum operator, $m_{\rm e}$ is the
effective electron mass, $\vec{\sigma}$ is the spin Pauli matrix, $\vec{m}$
is the direction of local magnetization, and $J$ is the exchange energy. In
Sec.~\ref{Sec:Derivation}, we show that in the slowly varying limit, the
system can be described by the \emph{locally defined} eigenstates which are
denoted by $|\vec{k},\pm\rangle_{(0)}$. Here $\vec{k}$ corresponds to the
electron momentum and $\pm$ is for minority and majority states. The
subscript $(0)$ refers to the eigenstates unperturbed by an electric field.
The eigenstates have spins aligned with the magnetization but with small
deviations as discussed in Refs.~\onlinecite{Xiao06PRB} and
\onlinecite{Aharonov92PRL} and illustrated in Fig.~\ref{Fig:Spin Spiral}(a).
The local spin expectation value for the unperturbed eigenstates is
\begin{equation}
\vec{\sigma}_{\vec{k},\pm}^{(0)}=\pm\vec{m}\mp\frac{\hbar}{2J}\vec{m}\times
(\vec{v}_{\vec{k}}\cdot\nabla)\vec{m},
\label{Eq:Spin density (unperturbed, lab)}
\end{equation}
where $\vec{v}_{\vec{k}}=\hbar\vec{k}/m_{\rm e}$ is the velocity of the
$|\vec{k},\pm\rangle_{(0)}$ state. In equilibrium, the deviations cancel on
summing up over all occupied states. However with non-equilibrium electron
distributions, they give rise to the \emph{current-induced} adiabatic spin
torque. If an electron relaxation mechanism is present, it relaxes the net
deviations, giving the \emph{current-induced} non-adiabatic spin
torque.\cite{Zhang04PRL}

When an electric field ${\bf E}$ is applied, it perturbs the eigenstates and
generates an additional deviation in the spin direction. With the perturbed
eigenstates,
$\vec{\sigma}_{\vec{k},\pm}=\vec{\sigma}_{\vec{k},\pm}^{(0)}+\Delta\vec{\sigma}_{\vec{k},\pm}$
where
\begin{equation}
\Delta\vec{\sigma}_{\vec{k},\pm}=\pm\frac{\hbar^2e}{4m_{\rm
    e}J^2}(\vec{E}\cdot\nabla)\vec{m}.
\label{Eq:Spin density (perturbed, lab)}
\end{equation}
Here $e>0$ is the electron charge. We demonstrate below that this deviation
in the spin direction gives an intrinsic contribution to the non-adiabatic
spin torque. Equation~(\ref{Eq:Spin density (perturbed, lab)}) is
\emph{electric-field-induced} and is a main result of this paper. This simple
picture for the origin of the torque is essentially the same as that
given\cite{Kurebayashi14NN} for the intrinsic spin-orbit torque, which is
also electric-field-induced, but differs from that given\cite{Gorini08PRB}
for the current-induced spin polarization, which is a current-induced effect,
based on its dependence on the momentum relaxation time. The perturbation due
to the electric field here has a characteristic length
$\Delta\vec{r}=\hbar^2e\vec{E}/4m_{\rm e}J^2$.  In Fig.~\ref{Fig:Spin
Spiral}(b), we show that one way to understand Eq.~(\ref{Eq:Spin density
(perturbed, lab)}) is to imagine that the electric field shifts the spins
spatially by an amount $\Delta\vec{r}$ as in
\begin{equation}
\vec{\sigma}_{\vec{k},\pm}[\vec{m}(\vec{r},t)]=\vec{\sigma}_{\vec{k},\pm}^{(0)}\left[\vec{m}\left(\vec{r}+\Delta\vec{r},t\right)\right].\label{Eq:Spin shift}
\end{equation}
Expanding the functional on the right hand side to lowest order in $\vec{E}$
gives Eq.~(\ref{Eq:Spin density (unperturbed, lab)}) and Eq.~(\ref{Eq:Spin
  density (perturbed, lab)}).

The equation of motion for the magnetization is given by the
Landau-Lifshitz-Gilbert equation including spin torque contributions,
\begin{equation}
\partial_t\vec{m}=-\gamma\vec{m}\times\vec{H}_{\rm
eff}+\alpha\vec{m}\times\partial_t\vec{m}+\vec{T},
\end{equation}
where $\vec{H}_{\rm eff}$ is the effective magnetic field and $\alpha$ is the
Gilbert damping parameter. The spin torque $\vec{T}$ is calculated from
$\vec{T}=(J\gamma/M_{\rm
s})\sum_{\vec{k},s}\vec{m}\times\vec{\sigma}_{\vec{k},s}f_{\vec{k},s}$, where
$\gamma$ is the gyromagnetic ratio, $M_{\rm s}$ is the saturation
magnetization, and $f_{\vec{k},s}$ is the electron distribution function.
After some algebra, Eqs.~(\ref{Eq:Spin density (unperturbed, lab)}) and
(\ref{Eq:Spin density (perturbed, lab)}) lead to
\begin{align}
\frac{\partial\vec{m}}{\partial t}&=-\gamma\vec{m}\times\vec{H}_{\rm eff}+\alpha\vec{m}\times\partial_t\vec{m}+\frac{\mu_{\rm B}}{eM_{\rm s}}(\vec{j}_{\rm s}\cdot\nabla)\vec{m}\nonumber\\
&\quad-\frac{\beta\mu_{\rm B}}{eM_{\rm s}}\vec{m}\times(\vec{j}_{\rm s}\cdot\nabla)\vec{m}-\frac{n_{\rm s}\mu_{\rm B}\hbar e}{2m_{\rm e}JM_{\rm s}}\vec{m}\times(\vec{E}\cdot\nabla)\vec{m},\label{Eq:LLG}
\end{align}
where $\mu_{\rm B}$ is the Bohr magneton, $\vec{j}_{\rm
s}=e\sum_{\vec{k},s}s\vec{v}_{\vec{k}}f_{\vec{k},s}$ is the spin-polarized
electrical current density, $n_{\rm s}=-\sum_{\vec{k},s}sf_{\vec{k},s}$ is
the spin-polarized density,\cite{comment:sign convention} and $\beta$ is the
non-adiabaticity parameter.\cite{Thiaville05EPL,Zhang04PRL} To obtain
Eq.~(\ref{Eq:LLG}), we implicitly assume the existence of impurity potential
in addition to Eq.~(\ref{Eq:s-d Hamiltonian}). The momentum relaxation due to
the impurity potential determines the current and the spin current
$\vec{j}_{\rm s}$ and its spin relaxation determines the second ($\alpha$)
and fourth ($\beta$) terms, \cite{Kambersky07PRB,Garate09PRB,Zhang04PRL}
which here we have added by hand. The last term is affected by the impurity
potential through vertex corrections, but we neglect those effects until
Sec.~\ref{Sec:Derivation-remarks_vertex}, since the qualitative features are
unchanged. The last three terms are the spin torques that result when an
electric field is applied. The first of these terms,  the adiabatic spin
torque, comes from the changes in the occupation of the electron states
removing the cancellation of terms from Eq.~(\ref{Eq:Spin density
(unperturbed, lab)}). Note that it is proportional to $\vec{j}_{\rm s}$ and
the coefficient of proportionality is independent of the electron
momentum-relaxation lifetime, making it current-induced. The next term, the
current-induced non-adiabatic spin torque, comes from extrinsic spin
relaxation mechanisms from the impurity potential (see Fig.~\ref{Fig:Spin
Spiral} for instance) and proportional to $\vec{j}_{\rm s}$ as well.

The last term in Eq.~(\ref{Eq:LLG}), the new result in this paper, is
proportional to $\vec{E}$ and the coefficient of proportionality is
independent of the electron momentum-relaxation lifetime, making the term
electric-field-induced. This term is the finite result that arises from
summing $\Delta\vec{\sigma}_{\vec{k},\pm}$ over the equilibrium Fermi sea and
is the central result of this paper. The occupation changes associated with a
finite charge current only make higher order corrections to the result.  In
Appendix~\ref{Sec(A):Fisher-Lee}, we discuss, in the context of the
Fisher-Lee theorem,\cite{Fisher81PRB,Baranger89PRB} how perturbations summed
over the whole Fermi sea are related to transport properties typically
derived from electronic properties just at the Fermi surface. Since $\vec{E}$
and $\vec{j}_{\rm s}$ are proportional in typical meterials, the
electric-field-induced spin torque is also proportional to
$\vec{m}\times(\vec{j}_{\rm s}\cdot\nabla)\vec{m}$, so that it gives another
contribution to the non-adiabatic spin torque. Hence the
electric-field-induced spin torque plays the same role in domain wall
motion as the current-induced non-adiabatic spin torque. See
Sec.~\ref{Sec:Discussion} for further discussion.

Although we demonstrate our theory for a free electron (quadratic)
dispersion, the calculation proceeds in a similar way for an arbitrary
dispersion $\varepsilon(\vec{k})$, with an intuitive way of generalization.
See Eqs.~(\ref{Eq:Spin density (unperturbed, lab)2}) and (\ref{Eq:Spin
density (perturbed, lab)2}) in Sec.~\ref{Sec:Derivation} for more
information.

\section{Theory\label{Sec:Derivation}}
In this section, we present our theory more in detail. We first present in
Sec.~\ref{Sec:Derivation-derivation} the derivation of Eqs.~(\ref{Eq:Spin
density (unperturbed, lab)}) and (\ref{Eq:Spin density (perturbed, lab)}) [or
Eqs.~(\ref{Eq:Spin density (unperturbed, lab)2}) and (\ref{Eq:Spin density
(perturbed, lab)2}) more generally]. In the rest of this section, we present
some remarks. Since the key results required for the discussions from
Sec.~\ref{Sec:Discussion} are already summarized in Sec.~\ref{Sec:Result},
readers who are less interested in the formal details can skip this section.

\subsection{Electric-field-induced spin density~\label{Sec:Derivation-derivation}}
We start from the following Hamiltonian with an arbitrary dispersion
$\varepsilon(\vec{k})$.
\begin{equation}
H=\varepsilon(\vec{k})+J\vec{\sigma}\cdot\vec{m}(\vec{r},t).
\end{equation}
Here $\vec{k}=\vec{p}/\hbar=-i\nabla_\vec{r}$ is still an operator. In this
theory, we take the slowly varying limit, by keeping only terms up to first
order in derivatives of magnetization. In this limit, it is useful to
transform the coordinate system in spin space to make the magnetic texture
uniform along $\vhat{z}$.\cite{Tatara08PR,Tatara97PRL} We use a unitary
transformation of the wavefunction $\psi$ to $U^\dagger\psi$ with
$U^\dagger=e^{i\theta\sigma_y/2}e^{i\phi\sigma_z/2}$, where
$\theta(\vec{r},t)$ and $\phi(\vec{r},t)$ are defined by
$\vec{m}(\vec{r},t)=(\sin\theta\cos\phi,\sin\theta\sin\phi,\cos\theta)$.
After the transformation, the Schr\"{o}dinger equation for $H$ becomes that
for $H'$ where
\begin{align}
H'&=\varepsilon(\vec{k}-i U^\dagger\nabla U)+J\sigma_z-i\hbar U^\dagger\partial_t U\nonumber\\
&=\varepsilon(\vec{k})+J\sigma_z-
\sum_{i=x,y,z}v_i(\vec{k})\vec{\sigma}\cdot\vec{A}_i-\vec{\sigma}\cdot\vec{A}_t,
\label{Eq:Hamiltonian slowly varying}
\end{align}
up to first order in gradients. Here
$\vec{v}(\vec{k})=(1/\hbar)\nabla_\vec{k}\varepsilon$ is the generalized
velocity for the dispersion $\varepsilon(\vec{k})$. The magnetic texture
becomes uniform and the effect of the original non-uniform texture is
contained in $\vec{A}_\mu$, which is defined through
$\vec{\sigma}\cdot\vec{A}_\mu=i\hbar U^\dagger\partial_\mu U$
($\mu=x,y,z,t$). Note that $\vec{A}_i$ ($i=x,y,z$) and $\vec{A}_t$ account
for spatial and temporal variation of $\vec{m}$ respectively. The third term
in Eq.~(\ref{Eq:Hamiltonian slowly varying}) acts as an effective spin-orbit
coupling, allowing us to apply the theory of intrinsic spin-orbit
torque.\cite{Kurebayashi14NN} In most of this paper, we neglect $\vec{A}_t$
since it gives rise to only small renormalization of parameters, as we
demonstrate in Sec.~\ref{Sec:Derivation-remarks_A_t}.

To find the \emph{locally defined} eigenstates within the slowly varying
approximation, we neglect the spatiotemporal variation of $\vec{A}_i$ since
it arises from the second order derivatives $\partial_\mu
\partial_i \vec{m}$. Then, Eq.~(\ref{Eq:Hamiltonian slowly varying}) has translation symmetry and
$\vec{k}$ is a good quantum number, thus it can be treated as a $c$-number.
Thus, the local eigenstates of Eq.~(\ref{Eq:Hamiltonian slowly varying}) are
given by $|\vec{k},\pm\rangle_{(0)}$ and the local spin expectation value
without an electric field is
\begin{align}
\vec{\sigma}_{\vec{k},\pm}^{(0)}(\vec{r})&=\langle\vec{k},\pm|_{(0)}U^\dagger\vec{\sigma}U|\vec{k},\pm\rangle_{(0)}\nonumber\\
&=\pm\vec{m}\mp\frac{\hbar}{2J}\vec{m}\times
[\vec{v}(\vec{k})\cdot\nabla]\vec{m},\label{Eq:Spin density (unperturbed, lab)2}
\end{align}
giving Eq.~(\ref{Eq:Spin density (unperturbed, lab)}) for a free electron
dispersion, for which $\varepsilon(\vec{k})=\hbar^2\vec{k}^2/2m_e$ and
$\vec{v}(\vec{k})=\vec{v}_\vec{k}=\hbar\vec{k}/m_e$.

When an electric field is applied, it perturbs the electronic states. The
perturbation is found by replacing $\vec{p}$ by $\vec{p}+e\vec{E}t$, after
which the effective spin-orbit coupling in Eq.~(\ref{Eq:Hamiltonian slowly
varying}) induces inter-band transitions between majority
$|\vec{k},-\rangle_{(0)}$ and minority states $|\vec{k},+\rangle_{(0)}$. For
a small $\vec{E}$, time-dependent perturbation theory with an adiabatically
turned-on electric field gives modified wavefunctions $|\vec{k},\pm\rangle$
and a modified local spin expectation value
$\vec{\sigma}_{\vec{k},\pm}(\vec{r})=\langle\vec{k},\pm|U^\dagger\vec{\sigma}U|\vec{k},\pm\rangle$,
giving Eq.~(\ref{Eq:Spin density (perturbed, lab)2}). An alternate approach
is the Kubo formalism,\cite{Kurebayashi14NN,Sinova04PRL} which we adopt here
because it provides a compact description. The Kubo formula gives the
statistical average of the non-equilibrium spin density
$\Delta\langle\vec{\sigma}\rangle$ in the steady state as
\begin{align}
\Delta\langle\vec{\sigma}\rangle&=-e~\mathrm{Im}\sum_{\vec{k},s\ne s'}\frac{f_{\vec{k},s}-f_{\vec{k},s'}}{(E_{\vec{k},s}-E_{\vec{k},s'})^2}\langle\vec{k},s|U^\dagger\vec{\sigma}U|\vec{k},s'\rangle\nonumber\\
&\quad\quad\quad\quad\quad\quad\times\langle\vec{k},s'|(\vec{E}\cdot\nabla_\vec{k})H'|\vec{k},s\rangle,\label{Eq:Kubo}
\end{align}
where $E_{\vec{k},s}$ is the local energy eigenvalue corresponding to
$|\vec{k},s\rangle$ state. Here $(\vec{E}\cdot\nabla_\vec{k})H'$ gives
the velocity operator along the electric field direction
multiplied by $\hbar$. Since the off-diagonal element of the velocity
operator in spin space is proportional to $\vec{A}_i$, one can neglect all
other $\vec{A}_i$ contributions in the slowly varying approximation. For
instance, $(E_{\vec{k},s}-E_{\vec{k},s'})^2=4J^2$. A straightforward
calculation gives
\begin{align}
\Delta\langle\vec{\sigma}\rangle&=\sum_{\vec{k},s}\Delta\vec{\sigma}_{\vec{k},s}f_{\vec{k},s},\label{Eq:Spin density (perturbed, lab)2 averaged}\\
\Delta\vec{\sigma}_{\vec{k},\pm}&=\pm\frac{\hbar^2e}{4J^2}\sum_{ij}E_i
[M^{-1}(\vec{k})]_{ij}\partial_j\vec{m},\label{Eq:Spin density (perturbed, lab)2}
\end{align}
with the generalized mass tensor $[M^{-1}(\vec{k})]_{ij}=(1/\hbar^2)
\partial^2\varepsilon/\partial k_i\partial k_j$. When the free electron
dispersion $\varepsilon=\hbar^2\vec{k}^2/2m_e$ is taken,
$[M^{-1}(\vec{k})]_{ij}=m_e^{-1}\delta_{ij}$ giving Eq.~(\ref{Eq:Spin density
(perturbed, lab)}). The arbitrariness of $f_{\vec{k},s}$ at this stage
indicates that
Eq.~(\ref{Eq:Spin density (perturbed, lab)2}) holds for each
$|\vec{k},s\rangle$ state.

A remark is in order. Equation~(\ref{Eq:Spin density (perturbed, lab)2
averaged}) gives no contribution for an insulator. Since Eq.~(\ref{Eq:Spin
density (perturbed, lab)2 averaged}) is an electric-field-induced
contribution, which does not depend on a change in occupation, it is not
obvious that the result is zero. However, it is straightforward to verify
that summing Eq.~(\ref{Eq:Spin density (perturbed, lab)2}) over a completely
filled band gives zero.

\subsection{Vertex corrections~\label{Sec:Derivation-remarks_vertex}}

Previous calculations of spin transport properties have highlighted
the importance of calculating beyond lowest order in perturbation
theory, in particular the necessity of including vertex corrections.
In general, non-equilibrium quantities calculated from the Kubo
formula are sensitive to the existence of an impurity potential.
Vertex corrections arise from the fact that, even when one take the limit in
which the impurity concentration goes to zero, it gives a finite correction
to the final result. The correction depends on the band structure of the
system and the detailed properties of the impurities.

The effects of vertex corrections have been intensively studied for the
intrinsic spin Hall conductivity for a two-dimensional Rashba
model.\cite{Sinova04PRL} In this section, we make a parallel argument to
demonstrate the significance of vertex corrections for various models. First,
the intrinsic spin Hall conductivity for a two-dimensional Rashba model is
exactly canceled by vertex corrections from nonmagnetic
impurities.\cite{Inoue03PRB,Inoue04PRB,Mishchenko04PRL,Chalaev05PRB,Raimondi05PRB,Khaetskii06PRL}
Even when magnetization is introduced, the intrinsic anomalous Hall
conductivity for the Rashba model\cite{Inoue06PRL} also suffers an exact
cancellation. However, exact cancellation only occurs in this specific model
and any differences from this model prevent exact
cancellation.\cite{Nomura05PRB,Murakami04PRB} A recent
experiment\cite{Kurebayashi14NN} on (Ga,Mn)As confirms the robust existence
of the intrinsic spin-orbit torque in real materials whose dispersion
deviates from a quadratic dispersion in the Rashba model. Moreover even for
the Rashba model, the existence of magnetic impurities changes the situation
drastically and vertex corrections may even enhance the intrinsic spin Hall
conductivity and intrinsic anomalous Hall conductivity.\cite{
Ren08JPC,Kato07NJP,Inoue06PRL,Moca07NJP,Gorini08PRB,Milletari08EPL}

The situation is similar for intrinsic spin torques as seen in the
mathematical structure of Eq.~(\ref{Eq:Hamiltonian slowly varying}), which is
the same as the two-dimensional Rashba model. We demonstrate in
Appendix~\ref{Sec(A):Chiral connection} that the Rashba Hamiltonian is a
special case of Eq.~(\ref{Eq:Hamiltonian slowly varying}) for a particular
magnetic texture. Therefore, we can adopt the results found for the Rashba
model.~\cite{Qaiumzadeh15PRB}  These results imply that for non-magnetic
impurities and a free electron band structure, vertex corrections exactly
cancel our main result. However, that cancellation only holds for that
particular model, for example Ref.~\onlinecite{Ren08JPC} gives the vertex
corrections for a magnetic impurity potential $V=\sum_i\int
d\vec{r}\delta(\vec{r}-\vec{R}_i)(\sigma_xS_x+\sigma_yS_y+\gamma
\sigma_zS_z)$, where $u$ characterizes the strength of the impurity
potential, $\vec{S}$ is the impurity spin with random direction, $0<\gamma<1$
is the anisotropy of the interaction, and $\vec{R}_i$ is the position of the
impurity. Equation~(29) in Ref.~\onlinecite{Ren08JPC} shows that the spin
Hall conductivity can be even enhanced by the factor
$1+2\gamma^2/(2+\gamma^2)$. This clearly shows that the intrinsic
non-adiabatic spin torque does not vanish due to vertex corrections unless
\emph{all} impurities are nonmagnetic.\cite{comment:vertex correction} In
fact, it can be even enhanced for some magnetic impurity potentials.

As for the Rashba model, when the dispersion deviates from strictly quadratic
behavior, there is no exact cancellation even if all impurities are
nonmagnetic. However, the situation is slightly different from the Rashba
model in our case. In our case, the form of effective spin-orbit coupling
also changes [See Eq.~(\ref{Eq:Hamiltonian slowly varying})] when the
dispersion changes. For example, the profile described in
Appendix~\ref{Sec(A):Chiral connection} gives an effective spin-orbit
coupling of the Rashba form, $H=H'+V$ where
\begin{equation}
H'=\varepsilon(\vec{k})+\alpha_{\rm R}[v_y(\vec{k})\sigma_x-v_x(\vec{k})\sigma_y]+J\sigma_z,\label{Eq:Rashba-like}
\end{equation}
with $\alpha_{\rm R}$ characterizing the rate of change of the magnetization.
Since we are interested in the slowly varying limit of the magnetization, we
keep only first order terms in $\alpha_{\rm R}$. The impurity potential $V$
satisfies $\langle V(\vec{r})V(\vec{r}')\rangle=n_iu^2$, where the bracket
means the ensemble average, $n_i$ is the impurity concentration, and $u$
characterizes the strength of the impurity potential. We assume that
$\epsilon(\vec{k})$ is an even function of $k_x$ and $k_y$. Then,
$v_x(\vec{k})$ and $v_y(\vec{k})$ are odd in $k_x$ and $k_y$ respectively.

We follow the procedure in Ref.~\onlinecite{Ren08JPC}. Let us consider the
case that an electric field is applied along $x$ direction. Then, in the Kubo
formula Eq.~(\ref{Eq:Kubo}), $\vec{E}\cdot\nabla_\vec{k} H'=\hbar
v_x(\vec{k})+\hbar\alpha_{\rm
R}[M_{yx}^{-1}(\vec{k})\sigma_x-M_{xx}^{-1}(\vec{k})\sigma_y]\equiv\hbar\nu_x(\vec{k})$.
Vertex corrections give corrections to the current vertex by
$\nu_x(\vec{k})+\tilde{v}_x$. The equation for the vertex corrections is
\begin{equation}
\tilde{v}_x=\frac{n_iu^2}{L^2}\sum_{\vec{k}}G^{\rm A}(E_{\rm F},\vec{k})(\nu_x+\tilde{v}_x)G^{\rm R}(E_{\rm F},\vec{k}),\label{Eq:vertex correction}
\end{equation}
where $L^2$ is the area of the two-dimensional system, $E_{\rm F}$ is the
Fermi level, and $G^{\rm R/A}$ are the retarded and advanced Green's
functions. The Green's functions are defined by $G^{\rm
R/A}(E,\vec{k})=[E-H'(\vec{k})-i\mathrm{Im}\Sigma^{\rm R/A}]^{-1}$ where the
self-energies $\Sigma^{\rm R/A}$ are given by $\Sigma^{\rm
R/A}(E)=(n_iu^2/L^2)\sum_\vec{k} [E-H'(\vec{k})\pm i\eta]^{-1}$.
Here $\eta$ is an infinitesimally small number. Thus the Sokhotski-Plemelj
identity gives $\mathrm{Im}(x\pm i\eta)^{-1}=\mp \pi\delta(x)$. By using
this, one can show that, up to $\mathcal{O}(\alpha_{\rm R})$,
\begin{equation}
\mathrm{Im}\Sigma^{\rm R/A}_{ss'}(E)=\mp \frac{\pi n_iu^2}{L^2}
D_s(E)\delta_{ss'},
\end{equation}
where $D_s(E)=\sum_\vec{k}\delta(E-E_{\vec{k},s})$ is the density of state
for each spin band.

Since all the expressions are diagonal in $\vec{k}$, the self-consistent
equation Eq.~(\ref{Eq:vertex correction}) is a $2\times2$ matrix equation
which is exactly solvable, even though it is complicated. The situation
becomes much simpler in the clean limit $n_i\to0$. Although the right-hand
side of Eq.~(\ref{Eq:vertex correction}) is proportional to $n_i$, there is a
finite contribution from $1/(x^2+n_i^2)\to(\pi/n_i)\delta(x)$ that cancels
the factor $n_i$ in general. Keeping such contributions gives the solution of
Eq.~(\ref{Eq:vertex correction}),\cite{comment:vertex correction solution}
\begin{equation}
\tilde{v}_x=-\sigma_y\frac{\alpha_{\rm R}}{2J}\sum_{\vec{k},s}\frac{s v_x(\vec{k})^2}{D_s(E_{\rm F})}\delta(E_{\rm F}-E_{\vec{k},s}).\label{Eq:vertex correction solution}
\end{equation}
When summed up over all $\vec{k}$, the parity characteristics of
$\vec{v}(\vec{k})$ and $M_{ij}^{-1}(\vec{k})$ give Eq.~(\ref{Eq:vertex
correction solution}). $v_i(\vec{k})$ is an odd function of $k_i$,
$M_{yx}^{-1}(\vec{k})$ is an odd function of both $k_x$ and $k_y$, and
$M_{xx}^{-1}(\vec{k})$ is an even function of both $k_x$ and $k_y$. These
relationships make many of the complicated terms zero after summation.

Equation~(\ref{Eq:vertex correction solution}) is in a simple form but not so
transparent. It can be made more transparent for the case of a circular
dispersion $\varepsilon(\vec{k})=\varepsilon(k)$ where $k=|\vec{k}|$, and
$|\vec{v}(\vec{k})|=(1/\hbar)\varepsilon'(k)\equiv v(k)$. The energy
eigenvalues are given by $E_s(k)\equiv E_{\vec{k},s}=\varepsilon(k)+sJ$, up
to $\mathcal{O}(\alpha_{\rm R})$. Without loss of generality, let
$\varepsilon(0)=0$. In this case, there is a single Fermi wave vector $k_{\rm
F,\it s}$ satisfying $E_{s}(k_{\rm F, \it s})=E_{\rm F}$. The summation can
be converted to an integration over the two-dimensional $\vec{k}$ space, and
the integration can be easily performed due to the delta function. As a
result, the vertex correction is
\begin{equation}
\tilde{v}_x=\sigma_y\frac{\alpha_{\rm R}}{4J}[v(k_{\rm F,-})^2-v(k_{\rm F,+})^2\Theta(E_{\rm F}-J)],
\end{equation}
where $\Theta$ is the Heaviside step function.

For a two-dimensional Rashba model with a free electron dispersion as an
example, $v(k_{\rm F,-})^2-v(k_{\rm F,+})^2=4J$ so that
$\tilde{v}_x=\alpha_{\rm R}\sigma_y$ cancels the spin-orbit coupling
contribution exactly when the both bands are occupied, $E_{\rm F}>J$.
However, such a cancelation is not general for arbitrary
dispersions. For example, if the dispersion takes the form of
\begin{equation}
\varepsilon=\left\{\begin{array}{cl}
              \epsilon_0(1-\cos k \chi) &~\mathrm{for}~k<\pi/2\chi,\\
              \epsilon_0(k\chi-\pi/2)+\epsilon_0 &~\mathrm{for}~k\ge\pi/2\chi,
            \end{array}\right.
\end{equation}
which is continuous and differentiable function (up to second order),
$v(k_{\rm F,-})=v(k_{\rm F,+})$ for $E_{\rm F}>J+\epsilon_0$ thus there is no
vertex correction for this regime. This example clearly shows that the exact
cancelation for a free electron dispersion is not general.

\subsection{Role of $\vec{A}_t$ : Renormalization of parameters~\label{Sec:Derivation-remarks_A_t}}

In this section, we briefly mention the role of $\vec{A}_t$ which we ignored.
Including $\vec{A}_t$, the same procedure leads to the
Landau-Lifshitz-Gilbert equation by
\begin{align}
\partial_t\vec{m}&=-\gamma'\vec{m}\times\vec{H}_{\rm eff}+\alpha'\vec{m}\times\partial_t\vec{m}+\frac{\mu_{\rm B}'}{eM_{\rm s}}(\vec{j}_{\rm s}\cdot\nabla)\vec{m}\nonumber\\
&\quad-\frac{\beta\mu_{\rm B}'}{eM_{\rm s}}\vec{m}\times(\vec{j}_{\rm s}\cdot\nabla)\vec{m}-\frac{n_{\rm s}'\mu_{\rm B}'\hbar e}{2m_e JM_{\rm s}}\vec{m}\times(\vec{E}\cdot\nabla)\vec{m},
\end{align}
where $\gamma'=\gamma/(1+n_{\rm s}\gamma\hbar/2M_{\rm s})$ and
$\alpha'=\alpha/(1+n_{\rm s}\gamma\hbar/2M_{\rm s})$ are respectively the
renormalized gyromagnetic ratio and the renormalized Gilbert damping
parameter, and $\mu_{\rm B}'=\gamma'\hbar/2$ is the renormalized Bohr
magneton. Note that taking into account $\vec{A}_t$ does not change the form
of the Landau-Lifshitz-Gilbert equation, but only renormalizes several
parameters. As demonstrated in Ref.~\onlinecite{Zhang04PRL}, the
renormalization is negligible, justifying neglecting $\vec{A}_t$.

\subsection{Quasi-steady state approximation and the conservation of angular momentum~\label{Sec:Derivation-remarks_angular momentum}}

In this section, we discuss a crucial yet implicit assumption of our
calculation.  We follow the standard approach for perturbative calculations
in which the perturbation gives transitions from initial states that are
eigenstates of the unperturbed Hamiltonian to final states that are as well.
This implicitly assumes that the density matrix before and after the
perturbation lacks coherence between these eigenstates.  This approach has
been justified by Redfield,\cite{Redfield57IBM} who showed that even very
weak coupling of the states to a random bath removes the coherence from the
density matrix.  In general, this assumption does not cause any concern and
deserve any extra discussion. In the present case, however, the loss of the
coherence plays an intriguing role with respect to the conservation of
angular momentum. So we discuss this point further.

As we describe in
Sec.~\ref{Sec:Derivation-derivation}, the spin eigenstates change when an
electric field is applied and the magnetization evolves. However, the changes
in the state do not necessarily imply that the statistical average of the
spin $\langle\vec{\sigma}\rangle=\mathrm{Tr}[\rho\vec{\sigma}]$ changes,
where $\rho$ is the density matrix. Although a new basis is formed at each
instantaneous time during magnetization dynamics, in general, the density
matrix written in the new basis will have off-diagonal components in the
spin. Without an additional angular momentum source, these off-diagonal
components cannot relax and the spin cannot change its value. In that case,
the spin system cannot reach steady state in the presence of an electric
field because there is nowhere for the angular momentum to go except back to
the magnetization. However, Redfield\cite{Redfield57IBM} demonstrated that a
density matrix for the spin system relaxes to a diagonal matrix in the
presence of a weak general coupling to a random bath (like a phonon bath).
This weak coupling allows for the transfer of angular momentum from the
conduction electrons to the lattice via the phonons provided the relaxation
process is fast compared to the magnetization dynamics. In transition metal
ferromangets, the magnetization dynamics is much slower than the electron
spin dynamics. Therefore, it is valid to assume that the electrons are in in
a quasi-steady state, in which case the density matrix can be treated as
diagonal at each instantaneous time. In this limit,
$\langle\vec{\sigma}\rangle=\sum_{\vec{k},s}\vec{\sigma}_{\vec{k},s}f_{\vec{k},s}$
justifying the formula for spin-transfer torque around Eq.~(\ref{Eq:LLG}) and
accounting for the angular momentum transfer.

A crucial point about this momentum transfer to the lattice caused by the
coupling of the spin system to the phonons, is that the size of the torque is
independent of the strength of this coupling, provided the coupling is not
too weak.  During the relaxation process, the random bath pushes angular
momentum to the lattice from the spin-magnetization system. The existence of
the lattice contribution to the angular momentum is crucial to provide a sink
for angular momentum. However, the amount of the angular momentum absorbtion
is determined by off-diagonal components of the density matrix, but not by
details of the relaxation process such as the relaxation rate. Therefore,
this spin-transfer torque does not depend on the relaxation rate, but depends
only on the existence of the relaxation process that brings the spin system
to steady state on a time scale fast compared to the magnetization dynamics.

Such a situation, in which a weak coupling plays a crucial role but does not
determine the size of the effect, is similar to the role of inelastic
scattering when the resistance of a material is dominated by impurity
scattering.  The inelastic scattering is crucial for the existence of a
steady state current flow but does not determine the resistance or even the
net rate of heat generation.  Similarly here, the weak coupling to the bath
is crucial for the flow of angular momentum to and from the bath but does not
determine the rate of the flow.

We emphasize that the assumptions made here hold very generally, particularly
in spintronics. This assumption seems more crucial for our case, since we do
not include any explicit spin-orbit coupling in the Hamiltonian, making it
straightforward to track the angular momentum flow. In other calculations,
the same assumptions are made, but the presence of a magnetic field or
spin-orbit coupling breaks angular momentum conservation for the
spin-magnetization subsystem, obscuring the importance of the assumptions.

\section{Discussion\label{Sec:Discussion}}

\subsection{Intrinsic non-adiabatic spin-transfer torque}
The last term in Eq.~(\ref{Eq:LLG}) from our theory gives an additional
contribution to the non-adiabatic spin-transfer torque, which we
refer to as ``intrinsic.'' In this section, we compare our result to the
current-induced contribution, which we refer to as ``extrinsic.'' To compare
these torques, we rewrite the intrinsic non-adiabatic spin torque using
$\vec{j}_{\rm e}=n_{\rm e}e^2\tau \vec{E}/m_{\rm e}$ in the Drude model. Here
$\vec{j}_{\rm e}$ is the charge current, $n_{\rm e}e^2\tau/m_{\rm e}$ is the
charge conductivity, $n_{\rm e}$ is the electron density, and $\tau$ is the
momentum-relaxation time. Assuming the current polarization is approximately
given by the electron polarization gives $\vec{j}_{\rm s}=(n_{\rm s}/n_{\rm
e})\vec{j}_{\rm e}$ and the intrinsic non-adiabatic spin torque is
$-\beta_{\rm int} (\mu_{\rm B}/e M_{\rm s}) \vec{m}\times(\vec{j}_{\rm
s}\cdot\nabla)\vec{m}$. The intrinsic non-adiabaticity $\beta_{\rm int}$ is
\begin{equation}
\beta_{\rm int}=\frac{\hbar}{2J\tau}.~\label{Eq:Beta_int}
\end{equation}
We compare $\beta_{\rm int}$ to $\beta$ in a similar model due to spin-flip
scattering,\cite{Zhang04PRL} for which $\beta$ is very similar to
Eq.~(\ref{Eq:Beta_int}).  There, $\beta=\hbar/2J\tau_{\rm sf}$ where
$\tau_{\rm sf}$ is the spin relaxation time rather than the momentum
relaxation time $\tau$. Note that $\tau$ is generally significantly smaller
than $\tau_{\rm
  sf}$. For typical parameters, $\tau=10^{-15}\rm~s$ to $10^{-14}\rm~s$ and
$J = 1\rm~eV$, one obtains $\beta_{\rm int}=0.03$ to $0.33$, which is
significantly larger than commonly reported values of $\beta\sim0.01$. In
fact, this comparison is a crude estimate of the order of magnitude because
$\beta_{\rm int}$ is sensitive to vertex corrections. To be more
quantitative, the vertex corrections discussed in
Sec.~\ref{Sec:Derivation-remarks_vertex} need to be taken into account.

The enhancement of $\beta$ due to the additional contribution $\beta_{\rm
int}$ leads to faster motion of magnetic domain
walls\cite{Thiaville05EPL,Zhang04PRL} and Skyrmion
lattices.\cite{Iwasaki13NN} For low currents, their velocity is proportional
to $\beta/\alpha$, where $\alpha$ is the damping parameter. Increasing the
extrinsic non-adiabaticity to increase this ratio is complicated by the fact
that the mechanisms that contribute to $\beta$ also contribute to
$\alpha$.\cite{Garate09PRB}  The ratio $\beta/\alpha$ tends to remain close
to one\cite{Duine07PRB,Barnes05PRL} even when the system is modified to
increase $\beta$. The intrinsic non-adiabaticity $\beta_{\rm int}$, on the
other hand, is not directly related to processes that contribute to $\alpha$.
$\alpha$ is defined as the damping rate for the precession of spatially
homogeneous $\vec{m}$. While true spin-orbit coupling contributes to
$\alpha$,\cite{Kambersky07PRB} the effective spin-orbit coupling in
Eq.~(\ref{Eq:Hamiltonian slowly varying}) is not a true spin-orbit coupling
and vanishes for spatially homogeneous $\vec{m}$.\cite{comment:no damping}
Thus, $\beta_{\rm int}/\alpha$ can be significantly larger than one.
Regarding experimental situations, there is no agreement on the ratio between
experimentally measured $\beta$ and $\alpha$: many experiments find the ratio
$\beta/\alpha$ to be close to one while some experiments\cite{Sekiguchi12PRL}
report large values for this ratio.  In those cases, $\beta_{\rm int}$ may be
playing a dominant role, which then suggests that it might be possible to
increase $\beta_{\rm int}$ while decreasing $\alpha$ to give more efficient
domain wall motion.

\subsection{Consistency with other theories\label{Sec:Consistency}}
In magnetization dynamics, many parameters that characterize the system are
not independent of each other; there are frequently close connections. A well
known such relationship is Onsager reciprocity. When a new contribution to
spin-transfer torque is discovered, its Onsager counterpart should be derived
in the same context, to be consistent. Another relationship is the chiral
connection\cite{Kim13PRL} we recently reported that gives a one-to-one
correspondence for each term appearing in the equations of motion for a
Rashba spin-orbit coupling system and those in a a textured magnetic system.
Thus, the intrinsic non-adiabatic spin-transfer torque is connected to a
contribution in a Rashba system.

\subsubsection{Onsager reciprocity\label{Sec:Onsager}}
The existence of the intrinsic non-adiabatic spin torque implies that there
is an additional contribution to the spin motive force $\vec{E}_{\pm}^{\rm
SMF}$\cite{Volovik87JPC,Barnes07PRL,Tserkovnyak02PRL} since they are related
by an Onsager relation. According to the Onsager relation, the intrinsic
non-adiabatic spin torque implies an intrinsic charge current $\vec{j}^{\rm
SMF}$ induced by the magnetization dynamics
where\cite{Tserkovnyak08PRB,Duine09PRB}
\begin{equation}
j_i^{\rm SMF}=\frac{n_{\rm s}e\hbar^2}{4m_{\rm e}J}\partial_i\vec{m}\cdot\partial_t\vec{m},~E_{\pm,i}^{\rm SMF}=\mp\frac{\hbar}{2e}\beta_{\rm
int}\partial_i\vec{m}\cdot\partial_t\vec{m}.\label{Eq:SMF}
\end{equation}
The left expression is the current predicted from the Onsager relation, and
the right expression is the spin-dependent electric field giving
$\vec{j}^{\rm SMF}$ within the Drude model.

We verify for a drifting spin spiral configuration given by
Eq.~(\ref{Eq(S):spiral profile}) that the inter-band transition contribution
due to magnetization dynamics\cite{Thouless83PRB} indeed generates such
charge current. The electrical current density $j_e$ due to inter-band
transitions is given by
\begin{align}
j_e&=-\frac{e\hbar^2}{2\pi m_{\rm e}}\int dk\frac{f_{k,-}(1-f_{k,+})}{E_{k,-}-E_{k,+}}\nonumber\\
&\quad\quad\quad\quad\quad\times\langle\partial_x\psi_{k-}|\psi_{k+}\rangle\langle \psi_{k+}|\partial_t\psi_{k-}\rangle+\mathrm{h.c.},
\end{align}
where $\psi_{ks}$ represents the instantaneous eigenstate neglecting
$\partial_t\vec{m}$, $s=\pm$ corresponds to minority and majority bands, and
h.c. refers to the hermitian conjugate. Here $k$ is a scalar since the system
is one-dimensional. $\partial_x\psi_{ks}$ and $\partial_t\psi_{ks}$
respectively come from current operator and $\partial_t\vec{m}$. Using the
eigenstates presented in Refs.~\onlinecite{Xiao06PRB,Calvo78PRB}, after some
algebra one obtains
\begin{equation}
\langle\psi_{k+}|\partial_\mu\psi_{k-}\rangle=-\langle\partial_\mu\psi_{k-}|\psi_{k+}\rangle=-\frac{i}{2}\partial_\mu\theta\cos\alpha_k.
\end{equation}
Keeping lowest order terms in derivatives, one can use
$E_{k,-}-E_{k,+}=-2J\equiv-\hbar^2 k_{\rm B}^2/m_{\rm e}$ and
$\cos\alpha_k=1$. Finally, using
\begin{equation}
\int dk f_{k-}(1-f_{k+})=2(\sqrt{k_{\rm F}^2+k_{\rm B}^2}-\sqrt{k_{\rm F}^2-k_{\rm B}^2}),
\end{equation}
where $k_{\rm F}$ is the Fermi wave vector, one obtains
\begin{equation}
j_e=(n_--n_+)\frac{e\hbar^2}{4m_{\rm e}J}\partial_x\theta\partial_t\theta,
\end{equation}
where $n_\pm=\sqrt{k_{\rm F}^2\mp k_{\rm B}^2}/\pi$ is the minority/majority
electron density. This expression is equivalent to Eq.~(\ref{Eq:SMF}). As we
see in Appendix~\ref{Sec(A):spin spiral}, inter-band transitions are captured
by considering $\vec{A}_t$ in our language. Thus, for the Onsager
counterpart, one should take into account $\vec{A}_t$ even though it gives
negligible effects for spin torques.

Equation~(\ref{Eq:SMF}) is of the same form as the non-adiabatic spin motive
force\cite{Duine09PRB,Tserkovnyak08PRB} but can be larger since $\beta_{\rm
int}$ can be larger than extrinsic contributions to $\beta$. In addition, its
chiral connection (See Sec.~\ref{Sec:Chiral connection}) gives a large
non-adiabatic spin-orbit motive force which can be larger than the extrinsic
contribution.\cite{Kim13PRL}

\subsubsection{Chiral connection to spin-orbit torques\label{Sec:Chiral connection}}

We have shown earlier\cite{Kim13PRL} that there is a one-to-one
correspondence between effects due to spatial variation of $\vec{m}$ and
those due to Rashba spin-orbit coupling, $(\alpha_{\rm
R}/\hbar)\vec{\sigma}\cdot(\vec{p}\times\vhat{z})$, where $\alpha_{\rm R}$ is
the Rashba parameter and $\vhat{z}$ is the surface normal direction. Rashba
spin-orbit coupling effects can be obtained by simply replacing conventional
derivatives $\partial_i\vec{m}$ by chiral derivatives
$\widetilde{\partial}_i\vec{m}=
\partial_i\vec{m}+k_{\rm R}(\vhat{z}\times\vhat{x}_i)\times\vec{m}$ in the equation of motion,
where $k_{\rm R}=2\alpha_{\rm R}m_{\rm e}/\hbar^2$ and $\vhat{x}_i$ is the
unit vector along $i$ direction.  This chiral derivative applied to the
magnetization texture follows from the covariant
derivatives\cite{Tokatly08PRL,Gorini10PRB} that have been applied to
electronic states and vector potentials in these same systems.

An example of this correspondence is between the interfacial
Dzyaloshinskii-Moriya interaction\cite{Moriya60PR,Dzyaloshinskii57SPJ} and
the micromagnetic exchange energy. Out of equilibrium, current-induced
field-like spin-orbit torques\cite{Obata08PRB,Matos09PRB,Manchon08PRB} and
damping-like spin-orbit torques\cite{Kim12PRB,Pesin12PRB,Wang12PRL}
correspond to current-induced adiabatic and nonadiabatic spin torques,
respectively. For the intrinsic non-adiabatic spin torque in
Eq.~(\ref{Eq:LLG}), replacing $\vec{m}\times(\vec{E}\cdot\nabla)\vec{m}$ by
the chiral derivative $\vec{m}\times(\vec{E}\cdot\widetilde{\nabla})\vec{m}$
generates the original term and an additional torque term,
\begin{equation}
\vec{T}_{\rm R}^{\rm int}=k_{\rm R}\frac{n_{\rm s}\mu_{\rm B}\hbar e}{2m_{\rm e}JM_{\rm s}}\vec{m}\times[\vec{m}\times(\vhat{z}\times\vec{E})],
\end{equation}
which is exactly the intrinsic spin-orbit torque reported in
Ref.~\onlinecite{Kurebayashi14NN} and which was calculated by a Berry phase.
The equivalence of these approaches can be verified by observing the relation
between the Kubo formula and the Berry phase.\cite{Thouless82PRL} In a
similar way, when combined with the intrinsic non-adiabatic spin torque, a
proper generalization of the chiral derivative provides an easy way to obtain
a Berry phase spin-orbit torque from other types of linear spin-orbit
coupling such as Dresselhaus spin-orbit couping\cite{Dresselhaus55PR} and
Weyl spin-orbit coupling.\cite{Anderson12PRL} We explicitly demonstrate in
Appendix~\ref{Sec(A):Chiral connection} that Rashba spin-orbit coupling and
Dresselhaus spin-orbit coupling are two particular cases.

\section{Summary\label{Sec:Summary}}

In summary, electric-field-induced changes in electronic states make an
intrinsic contribution to the non-adiabatic spin torque.  This contribution
arises from modifications to the states over the whole Fermi sea and is
independent of changes in the occupancy of the electron states.  Thus it
should be regarded as an electric-field-induced contribution rather than one
that is current-induced.  This effect, which occurs in the absence of
spin-orbit coupling, can be derived from a Berry phase due to the motion of
the electron spins through a spatially varying magnetization.  Through a
chiral connection, it is closely related to the intrinsic spin-orbit torque
that has been calculated from a Berry phase in a uniformly magnetized system
with Rashba
spin-orbit coupling.  While the magnitude of the intrinsic contribution is
sensitive to vertex corrections, we estimate that it is larger than other
contributions to the non-adiabatic spin torque at least in some systems.
Thus, it may play an important role in efficient electrical manipulation of
domain walls and Skyrmions.

\begin{acknowledgments}
KWK acknowledges stimulating discussions with D.~Go. KJL was supported by the
National Research Foundation of Korea (NRF) 
(2013R1A2A2A01013188). HWL and KWK were supported by NRF (2011-0030046,
2013R1A2A2A05006237) and the Ministry of Trade, Industry and Energy of Korea
(No.~10044723). KWK was supported by Center for Nanoscale Science and
Technology, National Institute of Standards and Technology, based on
Collaborative Research Agreement with Basic Science Research Institute,
Pohang University of Science and Technology. KWK was also supported by
Institute for Research in Electronics and Applied Physics at the University
of Maryland, based on Cooperative Research Program with the National
Institute of Standards and Technology.
\end{acknowledgments}

\begin{appendix}
\section{Spin expectation values for spin spirals~\label{Sec(A):spin spiral}}
\subsection{Drifting spin spiral}

The model is $\vec{m}=(\sin\theta\cos\phi,\sin\theta\sin\phi,\cos\theta)$
where
\begin{equation}
\theta(x,t)=px+\omega t,~\phi(x,t)=0.\label{Eq(S):spiral profile}
\end{equation}
Then, one immediately obtains from Eqs.~(\ref{Eq:Spin density (unperturbed,
lab)}) and (\ref{Eq:Spin density (perturbed, lab)})
\begin{equation}
\langle\vec{\sigma}\rangle_{\vec{k},\pm}=\pm\frac{J\vec{m}-\left(\frac{\hbar^2k_xp}{2m_{\rm e}}+\frac{\hbar\omega}{2}\right)\vhat{y}}{\sqrt{J^2+\left(\frac{\hbar^2k_xp}{2m_{\rm e}}+\frac{\hbar\omega}{2}\right)^2}}\pm\frac{\hbar^2e}{4m_{\rm e}J^2}E_x\partial_x\vec{m}.
\end{equation}
Here, $p$ comes from $\vec{A}_x$ and $\omega$ comes from $\vec{A}_t$. It is
illustrative to consider a few spacial cases.

\textbf{Case (i)} [$\omega=0$ and $E_x=0$].
\begin{equation}
\langle\vec{\sigma}\rangle_{\vec{k},\pm}=\pm\frac{J\vec{m}-\left(\frac{\hbar^2k_xp}{2m_{\rm e}}\right)\vhat{y}}{\sqrt{J^2+\left(\frac{\hbar^2k_xp}{2m_{\rm e}}\right)^2}}=\pm(\cos\alpha_{\vec{k}}\vec{m}-\sin\alpha_{\vec{k}}\vhat{y}),
\end{equation}
where
\begin{equation}
\sin\alpha_{\vec{k}}=\frac{k_xp}{k_x^2p^2+k_{\rm B}^2},~\frac{\hbar^2k_{\rm B}^2}{2m_{\rm e}}=J.
\end{equation}
This result agrees exactly with the result Eq.~(28) in
Ref.~\onlinecite{Xiao06PRB}. The physical implication of $\alpha_{\vec{k}}$
(or $\vec{A}_x$) is well discussed in the reference. $\alpha_{\vec{k}}$ is
shown in Fig.~\ref{Fig:Spin Spiral}(a) in the main text.

\textbf{Case (ii)} [$\omega\ne0$ and $E_x=0$].
\begin{align}
\langle\vec{\sigma}\rangle_{\vec{k},\pm}&=\pm\frac{J\vec{m}-\left(\frac{\hbar^2k_xp}{2m_{\rm e}}+\frac{\hbar\omega}{2}\right)\vhat{y}}{\sqrt{J^2+\left(\frac{\hbar^2k_xp}{2m_{\rm e}}+\frac{\hbar\omega}{2}\right)^2}}\nonumber\\
&=\pm(\cos(\alpha_{\vec{k}}+\varphi)\vec{m}-\sin(\alpha_{\vec{k}}+\varphi)\vhat{y}),\label{Eq(S):spin density (case 2)}
\end{align}
where
\begin{equation}
\sin\frac{\varphi}{2}=\frac{\hbar\omega/2}{\sqrt{(\hbar\omega/2)^2+J^2}}.
\end{equation}
There is an additional tilting towards $\vhat{y}$ direction by $\varphi$. One
finds a physical origin of $\varphi$ from inter-band transitions due to
$\partial_t\vec{m}$. Within the adiabatic approximation, the electronic
states can be approximated by the instantaneous eigenstates
$|\Psi\rangle\sim|\psi_0\rangle$ up to a phase factor. Considering the first
order inter-band transition, it reads\cite{Thouless83PRB}
\begin{equation}
|\Psi\rangle\approx e^{i\gamma_o(t)-\frac{i}{\hbar}\int^t dt'E_0(t')}\left[|\psi_0\rangle+i\hbar{\sum_{j\ne0}}|\psi_j\rangle\frac{\langle\psi_j|\partial_t|\psi_0\rangle}{E_j-E_0}\right],\label{Eq(S):inter band transition}
\end{equation}
with a Berry's phase $\gamma_j(t)=i\int^t
dt'\langle\psi_j|\partial_t|\psi_j\rangle$. One can show that the spin
expectation value from Eq.~(\ref{Eq(S):inter band transition}) is nothing but
Eq.~(\ref{Eq(S):spin density (case 2)}), implying that $\vec{A}_t$ captures
inter-band transitions during magnetization dynamics.

\textbf{Case (iii)} [$\omega\ne0$ and $E_x\ne0$].
\begin{align}
\langle\vec{\sigma}\rangle_{\vec{k},\pm}&=\pm\frac{J\vec{m}-\left(\frac{\hbar^2k_xp}{2m_{\rm e}}+\frac{\hbar\omega}{2}\right)\vhat{y}}{\sqrt{J^2+\left(\frac{\hbar^2k_xp}{2m_{\rm e}}+\frac{\hbar\omega}{2}\right)^2}}\pm\frac{\hbar^2e}{4m_{\rm e}J^2}E_x\partial_x\vec{m}\nonumber\\
&=
\pm\frac{J\vec{m}(x+\Delta x,t)-\left(\frac{\hbar^2k_xp}{2m_{\rm e}}+\frac{\hbar\omega}{2}\right)\vhat{y}}{\sqrt{J^2+\left(\frac{\hbar^2k_xp}{2m_{\rm e}}+\frac{\hbar\omega}{2}\right)^2}},\label{Eq:(S):spin shift}
\end{align}
where $\Delta x=\hbar^2eE_x/4m_{\rm e}J^2$. Note that Eq.~(\ref{Eq:(S):spin
shift}) differs from Eq.~(\ref{Eq(S):spin density (case 2)}) by changing the
argument $x$ of $\vec{m}$ to $x+\Delta x$. This is the spin shift discussed
in the main text.

\subsection{Rotating spin spiral}
The model is $\vec{m}=(\sin\theta\cos\phi,\sin\theta\sin\phi,\cos\theta)$
where
\begin{equation}
\theta(x,t)=px,~\phi(x,t)=\omega t.\label{Eq(S):spiral profile2}
\end{equation}
Then, one immediately obtains from Eqs.~(\ref{Eq:Spin density (unperturbed,
lab)}) and (\ref{Eq:Spin density (perturbed, lab)})
\begin{widetext}
\begin{equation}
\langle\vec{\sigma}\rangle_{\vec{k},\pm}=\pm\frac{J\vec{m}(x+\Delta x,t)-\frac{\hbar^2k_xp}{2m_{\rm e}}(\sin\omega t\vhat{x}-\cos\omega t\vhat{y})-\frac{\hbar\omega}{2}\vhat{z}}{\sqrt{J^2-J\hbar\omega\cos px+\frac{\hbar^2\omega^2}{4}+\frac{\hbar^4k_x^2p^2}{4m_{\rm e}^2}}}.
\end{equation}
\end{widetext}
For $\omega=0$ and $E_x=0$, the result is clearly consistent with
Ref.~\onlinecite{Xiao06PRB} as demonstrated in \textbf{[Case (i)]} for a
drifting spin spiral. In \textbf{[Case (ii)]} for a drifting spin spiral, for
non-zero $\omega$, inter-band transitions give rise to an additional tilting
angle $\varphi$. However, in this case the inter-band transitions do not give
rise to an additional tilting defined by a single value because
$\partial_x\vec{m}$ and $\partial_t\vec{m}$ are not parallel. One can still
observe that a finite $\omega$ gives rise to an additional tilting along
$\vhat{z}$ direction by the $-(\hbar\omega/2)\vhat{z}$ term. Also, it is
still clear that a spin shift with the same amount exists when an electric
field $E_x$ is applied as in \textbf{[Case (iii)]} as for a drifting spin
spiral.

\section{The Fisher-Lee theorem and its application to spin transfer
torques~\label{Sec(A):Fisher-Lee}}

It is appropriate to consider whether contributions summed over the whole
Fermi sea can affect transport properties.  The Fisher-Lee
theorem\cite{Fisher81PRB} and its multi-lead and magnetic field
generalization given by Baranger and Stone\cite{Baranger89PRB} state that in
a mesoscopic system, the conductivity can be determined purely from the
states at the Fermi energy.  A naive application of this theorem would
suggest that the effect described in this paper, built from contributions
from the whole Fermi sea, must be wrong. However, not only do these theorems
not directly apply to the situation under consideration, they in fact provide
support for our approach. These theorems apply to charges and to our
knowledge have not been successfully generalized to spin currents.  Further
they apply to the current and voltages going in and leaving a sample rather
than internal magnetization dynamics.  Nonetheless, the application of the
Baranger-Stone result to the anomalous Hall effect provides support for the
idea that the applied electric field affects the states over the whole Fermi
sea and that the effect can in turn affect the charge current.  There is a
large literature of the intrinsic or Berry-phase contribution to the
anomalous Hall conductivity, see Ref.~\onlinecite{Nagaosa10RMP} and
references therein.  This contribution is analogous to our result.  It arises
from the distortion of the wave functions by the electric field.  Naively
applied, the Fisher-Lee theorem would suggest that it must also be zero.
However, Sec.~VI~B in Ref.~\onlinecite{Baranger89PRB}, which discusses the
Fisher-Lee theorem as applied to the quantum Hall effect shows why it is not
zero.  The contributions to the quantum Hall conductivity calculated for a
bulk get modified by the edges of the sample.  In that case, the confining
potential pushes the Landau level states that are well below the Fermi level
in the bulk to the Fermi level at the edge, giving rise to the famous edge
states. There is a large literature on intrinsic effects for the anomalous
Hall effect, the spin Hall effect, and more recently spin-orbit torques,
which provided the inspiration of this work.  For these cases, the effect of
the spin-orbit coupling on the states well below the Fermi energy get pushed
to the Fermi energy near the edge of the sample.  In the present case, the
consequences of the effective spin-orbit coupling due to the magnetic texture
get pushed to the Fermi energy at the edges of the sample.

\section{Relation to Rashba and Dresselhaus spin-orbit couplings~\label{Sec(A):Chiral connection}}
In this section, we show that the Rashba and Dresselhaus spin-orbit couplings
are nothing but two particular cases of our theory within the first order
approximation. Here, one should note that it shows a mathematical equivalence
but not a physical equivalence of each system.

\subsection{Rashba model as a particular case\label{Section:Rashba}}
Consider an extremely slowly varying magnetic structure
$\vec{m}=(\sin\theta\cos\phi,\sin\theta\sin\phi,\cos\theta)$ as
\begin{equation}
\theta=\frac{\pi}{2}+px,~\phi=py,
\end{equation}
where the small parameter $p$ satisfies $pL\ll1$ for the system size $L$.
Then, one obtains up to $\mathcal{O}(p)$
\begin{equation}
\vec{A}_x=\frac{p\hbar}{2}\vhat{y},~\vec{A}_y=-\frac{p\hbar}{2}\vhat{x}.
\end{equation}
Then, the effective Hamiltonian within our theory reads
\begin{equation}
H_0'(\vec{k})=\frac{\hbar^2\vec{k}^2}{2m_{\rm e}}+\frac{p\hbar^2}{2m_{\rm e}}(\sigma_xk_y-\sigma_yk_x)+J\sigma_z,
\end{equation}
which is nothing but a Rashba model $H_{\rm SO}=\alpha_{\rm
R}\vec{\sigma}\cdot(\vec{k}\times\vhat{z})$ for $\alpha_{\rm
R}=p\hbar^2/2m_{\rm e}$.

\subsection{Dresselhaus model as a particular case}
Let
\begin{equation}
\theta=\frac{\pi}{2}+py,~\phi=px,
\end{equation}
for the same condition. Then, one obtains
\begin{equation}
\vec{A}_x=-\frac{p\hbar}{2}\vhat{x},~\vec{A}_y=\frac{p\hbar}{2}\vhat{y}.
\end{equation}
Now, the effective Hamiltonian within our theory reads
\begin{equation}
H_0'(\vec{k})=\frac{\hbar^2\vec{k}^2}{2m_{\rm e}}+\frac{p\hbar^2}{2m_{\rm e}}(\sigma_xk_x-\sigma_yk_y)+J\sigma_z,
\end{equation}
which is nothing but a Dresselhaus model $H_{\rm SO}=\alpha_{\rm
D}(\sigma_xk_x-\sigma_yk_y)$ for $\alpha_{\rm D}=p\hbar^2/2m_{\rm e}$.

\end{appendix}


\begin{thebibliography}{99}
\bibitem{comment:sign convention}%
Note that additional minus sign appears for spin quantities since $s=-1$
amounts to majority electrons in our notation. Note also that $n_{\rm s}$ is
the spin-polarized density not at the Fermi surface but that of the entire
Fermi sea.
\bibitem{comment:vertex correction}%
Effects of perfectly nonmagnetic impurities are addressed in the Appendix~A
of Ref.~\onlinecite{Tatara07JPSJ}, where all contributions to spin torque
from the Fermi sea (terms $S_2$, $S_4$, and $S_5$) are found to vanish in the
weak scattering limit.
\bibitem{comment:vertex correction solution}%
In fact, Eq.~(\ref{Eq:vertex correction}) allows the general solution to have
arbitrary diagonal components in spin space, but they are discarded since
there is no physical reason to have diagonal vertex corrections for our model
Eq.~(\ref{Eq:Rashba-like}). Also, they do not contribute to the final result
when put into the Kubo formula Eq.~(\ref{Eq:Kubo}).
\bibitem{comment:no damping}%
For non-homogeneous textures, there exists additional damping from the
effective spin-orbit coupling. However, it gives a higher order contribution
in derivative $\partial_\mu\partial_\nu\vec{m}$.
\end{thebibliography}
\end{document}